\documentclass[12pt]{article}
\usepackage{latexsym} \usepackage{epsfig,amssymb,euscript}
\usepackage{amsmath}

\def\be{\begin{equation}}
\def\ee{\end{equation}}
\def\bea{\begin{eqnarray}}
\def\eea{\end{eqnarray}}

\newcommand{\bZ}{\mathbb{Z}}

\newcommand{\non}{\nonumber}

\input epsf

\begin{document}

\pagestyle{empty}
%\vspace*{1.0in}
\rightline{SISSA-28/2009/EP}
\vspace{0.8cm}
\begin{center}
{\LARGE{\bf Natural semi-direct gauge mediation

\vskip 5pt
and D-branes at singularities \\ [10mm]}}

 {\large{Riccardo Argurio$^{1}$,
Matteo Bertolini$^{2}$, Gabriele Ferretti$^{3}$ \vskip 3pt and  Alberto Mariotti$^{4}$ \\[5mm]}}

{\small{{}$^1$ Physique Th\'eorique et Math\'ematique and International Solvay
Institutes \\
\vspace*{-2pt}  Universit\'e Libre de Bruxelles, C.P. 231, 1050
Bruxelles, Belgium\\

\medskip
{}$^2$ SISSA and INFN - Sezione di Trieste\\
\vspace*{-2pt} Via Beirut 2; I 34014 Trieste, Italy\\

\medskip
{}$^3$  Department of Fundamental Physics \\
\vspace*{-2pt} Chalmers University of Technology, 412 96 G\"oteborg, Sweden
\\

\medskip
{}$^4$ Theoretische Natuurkunde and International Solvay Institutes \\
\vspace*{-2pt}
Vrije Universiteit Brussel, Pleinlaan 2, B-1050 Brussels, Belgium}}\\

\medskip

\medskip

\medskip

\medskip

{\bf Abstract}
\vskip 20pt
%\end{center}
%\begin{center}
\begin{minipage}[h]{16.0cm}

We consider semi-direct gauge mediation models
of supersymmetry breaking where the
messengers are composite fields and their supersymmetric mass is
naturally generated through quartic superpotential couplings. We show
that such composite messenger models can be easily embedded in quiver gauge
theories arising from D-branes at Calabi-Yau singularities, and argue that
semi-direct gauge mediation is in fact a very natural option for supersymmetry
breaking in D-brane models. We provide several explicit examples and
discuss their salient phenomenological properties.

\end{minipage}
\end{center}
\newpage
%----------------------------------------------------------------------%
%  Resetting of counters
%----------------------------------------------------------------------%
\setcounter{page}{1} \pagestyle{plain}
\renewcommand{\thefootnote}{\arabic{footnote}} \setcounter{footnote}{0}

%%%%%%%%%%%%%%%%%%%%%%%%%%%%%%%%%%%%%%%%%%%%%%%%%%%%%%%%%%%%%%%%%%%%
%%%%%%%%%%%%%%%%%%%%%%%%%%%%%%%%%%%%%%%%%%%%%%%%%%%%%%%%%%%%%%%%%%%%
\section{Introduction}

Gauge mediation~\cite{original} (for a review see~\cite{Giudice:1998bp}) is one of the most popular
frameworks for a phenomenologically viable mechanism to transmit supersymmetry
breaking down to the MSSM.

Within the general messenger paradigm~\cite{messengers,Dine:1995ag}, a possibility which has been less investigated
so far is the one named ``semi-direct gauge mediation'' in~\cite{Seiberg:2008qj}. 
In this scheme of gauge mediation
the messengers interact with the hidden sector through gauge interactions
only, but, unlike in direct gauge mediation
models~\cite{direct}, they are irrelevant to the mechanism
of dynamical supersymmetry breaking (DSB). The only superpotential term involving the messengers is a mass
term. 
The hidden sector gauge group that
the messengers couple to (henceforth the ``messenger gauge group") can be
a (weakly gauged) global symmetry group, as in the mediator models of \cite{Randall:1996zi}, or
a genuine gauge group of the hidden sector, as for the models recently studied
in \cite{Seiberg:2008qj,Elvang:2009gk}. In the former case, sending to zero the
corresponding gauge coupling does not destroy supersymmetry breaking in the hidden sector.
In the latter case, having a non-vanishing gauge coupling is instead
crucial for the DSB mechanism to hold.

Semi-direct gauge mediation has some distinctive features as compared
to other popular frameworks of gauge mediation. Unlike minimal gauge mediation there
is no need for unnatural spurion-like superpotential couplings involving
singlets since
the messengers communicate with the hidden sector via gauge interactions
only. On the other hand, in
contrast to direct gauge mediation, the MSSM gauge group does not need to
be part of the flavor group of the DSB sector: the messenger gauge group
can be as small as $U(1)$ or $SU(2)$. This is sensibly softening the
problem of dangerous Landau poles due to the fact that DSB models with large flavor groups
tend to also have large gauge groups and thus a large number of messengers. 
Finally, in some specific
situations, semi-direct mediation models give rise to D-term like contributions to diagonal messenger masses, 
even in the absence of abelian gauge factors and FI
D-terms~\cite{Seiberg:2008qj,Elvang:2009gk}.

In semi-direct gauge mediation, the messengers, which carry MSSM quantum numbers, are usually considered
elementary fields, and their supersymmetric mass is a parameter of the theory. It
has to be at least some orders of magnitude below the Planck scale if gravity mediation effects
are to be suppressed. It is hence desirable to have a naturally small messenger
mass in these models.

In this note we first show that it is in fact not difficult to construct models of semi-direct gauge
mediation where all scales are dynamically generated, including the messenger masses
(we could refer to this as retrofitted \cite{Dine:2006gm}, or natural, semi-direct mediation), and where
the messengers themselves are composite fields. In direct mediation
models the messengers, as seen by the MSSM, are also naturally
composite fields. However in such models they are composite because of
strongly coupled dynamics in the DSB sector. The basic mechanism we present, which is similar in spirit to
that of \cite{Csaki:1997if}, relies instead on an additional gauge group, which we
dub ``mediator gauge group'', whose strong dynamics leads to the generation of composite
messengers, charged both under the hidden sector and
the visible sector gauge groups,
with a natural supersymmetric mass term which is independent on the details of the
hidden sector dynamics.

A second goal of this note is to show how such a framework arises quite naturally in
string theory. It is straightforward to picture the models above as quiver gauge
theories, with several gauge groups represented as nodes and bi-fundamentals going from one node
to another. This is the typical structure of the gauge theories describing D-branes
at Calabi-Yau (CY) singularities, suggesting that natural semi-direct mediation is rather
generic in string theory models of gauge mediation. To support such a claim, we will work out
several explicit examples where this holds.

The paper is organized as follows. In section 2 we describe the basic building blocks which are needed for
constructing semi-direct gauge mediation models with natural masses for the messengers. Using these building blocks,
in section 3 we discuss complete, explicit models of semi-direct gauge mediation. In section 4 we show how such
models arise quite generically in string theory, provide a few explicit examples, and discuss some of their very basic
phenomenological properties. 
Section 5 contains our conclusions and outlook.

%%%%%%%%%%%%%%%%%%%%%%%%%%%%%%%%%%%%%%%%%%%%%%%%%%%%%%%%%%%%%%%%%%%%%
%%%%%%%%%%%%%%%%%%%%%%%%%%%%%%%%%%%%%%%%%%%%%%%%%%%%%%%%%%%%%%%%%%%%%

\section{Retrofitting the messenger sector}

In what follows we will first describe the basic ingredients we are going to
use to generate a natural, composite messenger sector in the framework
of semi-direct gauge mediation. Then we will discuss how such basic building blocks
can be embedded in concrete gauge mediation models.

%%%%%%%%%%%%%%%%%%%%%%%%%%%%%%%%%%%%%%%%%%%%%%%%%%%%%%%%%%%%%%%%%%%%
\subsection{A toy model}
\label{section:toy}

\begin{figure}[ht]
\centering
%\hspace{1cm}
\includegraphics[width=.5\textwidth]{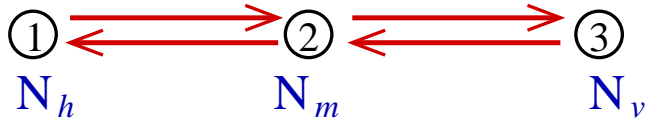}
\caption{\small The basic building blocks. Each node corresponds to a
$SU$ factor; the gauge group ranks are $N_h$, $N_m$ and $N_v$. The red
arrows correspond to bi-fundamental chiral superfields. We choose $N_m=N_h+N_v$.
\label{quivertoy}}
\end{figure}

Let us consider the non-chiral supersymmetric gauge theory depicted in
figure \ref{quivertoy}, a quiver gauge theory with three gauge factors
and bi-fundamental matter superfields. We will label the three gauge groups
as $SU(N_h)$ for the hidden sector gauge group (which, in more elaborate models, need not
be the gauge group whose strong dynamics leads to DSB), $SU(N_v)$
for the visible sector gauge group (typically, we will take
$N_v=5$), and $SU(N_m)$ for what we dubbed mediator gauge group. Of course,
generalizations to gauge groups other than $SU(N)$ are also possible.

An important further ingredient is to let
the bi-fundamental matter interact via a quartic superpotential:
\begin{equation}
\label{wtree}
W_{\mbox{\scriptsize tree}} = h_1 X_{hm} X_{mv} X_{vm} X_{mh} + h_2
X_{hm} X_{mh} X_{hm} X_{mh} + h_3 X_{vm} X_{mv} X_{vm} X_{mv}~.
\end{equation}
Traces over the indices of the various gauge groups are understood.
The couplings $h_i$ have dimension of an inverse mass and they are
inversely proportional to the UV scale generating the
non-renormalizable interaction (\ref{wtree}). In this toy model, which
lacks a UV completion, this scale is undetermined, and can be taken to
be the Planck scale.\footnote{In string theory embeddings, one can set
$h_i \sim 1/M^*_s$, with $M_s^*$ being the
string scale possibly warped down to a lower value by a duality cascade
RG flow. In order to avoid Landau pole problems, one might not like to have
$M_s^*$ too low, though. For definitess, in this paper we will always
take $M^*_s$ to be order of the Planck scale $M_p$.}

We now take the rank of the mediator gauge group to be
\be
N_m = N_h + N_v~.
\ee
Given this matter content, the beta function of the middle node will
be the one with the largest one-loop coefficient.
It is thus natural to assume that such gauge group reaches
strong coupling first. In other words, if
all the three gauge factors are UV free,
we assume the following hierarchy between the corresponding dynamically generated scales:
$\Lambda_m \gg  \Lambda_h,\Lambda_v$.
Since we want to achieve DSB we will need to add extra matter to the hidden gauge group.
Thus, we can also keep the possibility that the hidden gauge group is not
UV free, in which case we only impose  $\Lambda_m \gg  \Lambda_v$.

At scales above $\Lambda_m$ the other nodes act effectively
as global flavor groups and the theory reduces to SQCD with $N_f=N_c$
(where $N_c=N_m=N_h+N_v$ in this case).
The effective superpotential hence reads \cite{Seiberg:1994bz}
\begin{equation}
W_{\mbox{\scriptsize eff}} = h_1 M_{hv} M_{vh} + h_2 M_{hh} M_{hh} +
h_3 M_{vv} M_{vv} + \lambda \left(\mbox{det}\,{\cal M} - {\cal B}
\tilde{\cal B} - \Lambda_m^{2N_m}\right)~,
\end{equation}
with $\lambda$ a Lagrange multiplier, ${\cal B} \sim
(X_{mh})^{N_h}(X_{mv})^{N_v}$, $\tilde {\cal B} \sim
(X_{hm})^{N_h}(X_{vm})^{N_v}$ the baryon fields and ${\cal M} $ the meson
matrix, with entries $M_{hv}=X_{hm}X_{mv}$, $M_{vh}=X_{vm}X_{mh}$,
$M_{hh}=X_{hm}X_{mh}$ and $M_{vv}=X_{vm}X_{mv}$ (the summation
on the middle node gauge indices is understood).

Because of the quartic superpotential terms, which become quadratic
terms in the mesons, the moduli space separates into two disconnected
branches, the mesonic one and the baryonic one. On the mesonic branch
some mesons acquire VEVs, and hence the two gauge groups at the first
and third node are higgsed in some way. We will not be interested in
the physics along this branch.

Along the baryonic branch the low energy theory reduces to two 
SYM nodes ($SU(N_h)$
and $SU(N_v)$, respectively) coupled via the meson fields $M_{hv}$ and $M_{vh}$, and two
adjoints $M_{hh}$ and $M_{vv}$. The effective superpotential reduces to
\begin{equation}
W = h_1 M_{hv} M_{vh} + h_2 M_{hh} M_{hh} + h_3 M_{vv} M_{vv}~,
\end{equation}
which makes the (messenger!) fields $M_{hv}, M_{vh}$ and the two adjoints
$M_{hh}, M_{vv}$ massive, with a dynamically generated
(supersymmetric) mass
\begin{equation}
\label{messmass}
m_i  \sim h_i (\Lambda_m)^2~.
\end{equation}
Notice that we have to rescale the meson fields by $\Lambda_m$ to
reinstate the correct mass dimension, and we have assumed a canonical
K\"ahler potential near the
origin of the baryonic branch.  A hierarchy between the messenger and
the adjoint masses can be achieved either by hand by tuning the $h_i$,
or dynamically in slightly more elaborate models as we will show
later.\footnote{In this respect, note that the adjoint field $M_{vv}$ could
  play the role of the GUT Higgs field. This would indeed necessitate
some hierarchy between its mass scale and the scale of the messenger mass. Of
course, in order to implement this option in any specific model, one should
also find a mechanism to generate the other couplings of the GUT Higgs field.
The other option is that $M_{vv}$ is just a spectator. For a natural choice of
its mass scale it should not by itself induce a Landau pole.}

We note that models of semi-direct mediation where the messenger masses are generated
dynamically were also considered in \cite{Csaki:1997if}. There, however,
the effective dynamics of the mediator node was SQCD with a symplectic
gauge group and $N_f=N_c+1$ flavors \cite{Intriligator:1995ne}.

The simple mechanism described above may seem quite ad hoc. In fact,
this is not the case. A crucial ingredient for it to work is a  gauge
theory with a quiver structure and with quartic couplings between the
bi-fundamental chiral superfields. This is precisely what
happens, generically, when engineering supersymmetric gauge theories by means of
D-branes at singularities. In \S \ref{string} we will consider several such examples where
simple variants of the above mechanism arise automatically
when considering suitable stacks of fractional branes at CY singularities.

Let us end this section with a few more comments.

In the model described above, there is an additional massless field
given by the baryonic superfields modulo the baryonic branch constraint
${\cal B} \tilde{\cal B} = -\Lambda_m^{2N_m}$. This mode decouples completely
from the dynamics of the other nodes, and can be given a mass by gauging
the baryonic $U(1)$ global symmetry of the middle node
(in string theory setups,
this is naturally accomplished by compactification).

One more comment is about the absence of explicit mass terms in the tree-level superpotential
\eqref{wtree}. Such terms  are not forbidden by the gauge symmetries. In
order to prevent their appearance (which would lead at the effective
level to phenomenologically dangerous VEVs for the messengers), we could resort to a $\bZ_4$ symmetry acting
on the bi-fundamentals, or a continuous $U(1)_R$ symmetry. It is the
latter which forbids the mass terms when such quiver gauge theories
arise in string theory (typically, one defines the theory at a
superconformal fixed point, hence masses are obviously forbidden).
Another option is to have a slightly more complicated, but chiral,
model, as we will discuss later.

One could wonder whether models with a composite messenger sector generated with
a mechanism as the one described here may be plagued by possible instabilities.
Indeed, there are presumably supersymmetric states along the mesonic branch which survive even
in the regime where there is supersymmetry breaking on the baryonic branch. We will assume
that they are far enough in field space, typically as $\Delta X \sim \Lambda_m$. The
potential in the supersymmetry breaking vacuum is on the other hand
controlled by, say, $V \sim \Lambda_h^4$ so that tunnelling to the
supersymmetric vacua is parametrically suppressed by powers of
$\Lambda_h/\Lambda_m$.

It is obvious that the above simple toy model can be generalized in a
variety of ways. The main message we want to convey here is that it is
quite simple to build a supersymmetric gauge theory which, in some
regime, reduces to an effective theory with massive fields
transforming in the bi-fundamental representation of two otherwise
decoupled (gauge) sectors. For this to happen, quartic superpotential
terms are generically needed. The latter is  a nice feature of these
models since quartic terms are the most generic non-renormalizable
terms one can start adding to a given model. And, as already mentioned, such couplings
are ubiquitous in D-brane constructions of supersymmetric gauge theories.

%%%%%%%%%%%%%%%%%%%%%%%%%%%%%%%%%%%%%%%%%%%%%%%%%%%%%%%%%%%%%%%%%%%%
\subsection{More complete models}

\begin{figure}[ht]
\centering
%\hspace{1cm}
\includegraphics[width=.7\textwidth]{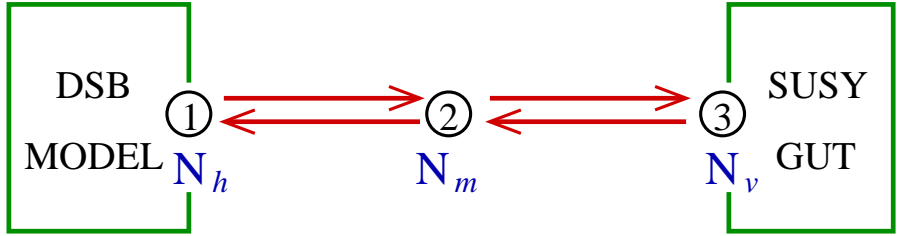}
\caption{\small The basic building blocks embedded into
semi-direct gauge mediation.
\label{sdgm}}
\end{figure}

This simple way of generating composite and naturally massive chiral
superfields mediating the interactions between two otherwise decoupled
sectors, can be easily embedded into concrete
semi-direct gauge mediation models.  This can be done by promoting
the hidden and visible nodes to full-fledged hidden and visible sectors,
including thus MSSM matter families and higgses on the visible side, and
any matter fields and additional gauge groups on the hidden side in order
to achieve DSB.
This is schematically
depicted in figure \ref{sdgm}.  This way, we have a set of (composite)
messenger fields, with a dynamically generated mass, coupled to the
hidden sector only through gauge interactions.

At scales well below $\Lambda_m$, the messengers behave as effectively
elementary. Thus,
regardless the specific supersymmetry breaking mechanism in the hidden
sector, the next step of the story amounts to computing the induced
non-supersymmetric messenger masses. Indeed, via gauge interactions with the gauge group $SU(N_h)$,
the messengers acquire a non-supersymmetric mass matrix which has
both diagonal and off-diagonal contributions. Such contributions are generically at
two-loops in the messenger group coupling and lead to a non-vanishing supertrace $\mbox{Str} M^2$ of the messenger mass matrix.

In fact, some interesting phenomena occur in specific situations. For instance, when
the messenger gauge group $SU(N_h)$ is higgsed due to the DSB
mechanism~\cite{Seiberg:2008qj,Elvang:2009gk}
there are also D-term-like contributions to the diagonal messenger masses which are induced by the F-terms.
Notably, these terms arise also in the absence of $U(1)$ factors (i.e. one does not need explicit
FI-terms, as in minimal gauge mediation). A second interesting phenomenon, which holds when the scale
of higgsing is higher than the messenger supersymmetric mass scale, is that the loop orders in
$SU(N_h)$ of the messenger masses are effectively reduced by one unit. This applies to the D-terms,
which appear then as classical, and to the aforementioned diagonal and off-diagonal contributions
which are now one-loop. The D-terms do not contribute to the supertrace of the
messenger sector masses, and the latter is thus one loop.
The off-diagonal masses (which are encoded as $F$ in minimal gauge mediation
models) depend differently on the scales of the model, and hence can
be naturally chosen to dominate or not.

One needs to worry about the messenger sector not having tachyonic
components. This is usually achieved by implementing a hierarchy among the
different scales of the problem. In our models it will be sufficient to
have a reasonable hierarchy among the dynamical scales of the hidden and the
mediator gauge groups. Very schematically, and in the worst case scenario,
one has to impose a relation like
\begin{equation}
\frac{m_{\mbox{\scriptsize non-susy}}}{m_\mathrm{susy} }
\sim  \frac {\Lambda_h}{h
(\Lambda_m)^2}~ \sim  \frac {\Lambda_h M_p}{
(\Lambda_m)^2}~ < 1
\end{equation}
Of course, such bounds will have to be refined in specific models,
in cases for instance where the hidden sector has more than one scale.
One has then more options when choosing a hierarchy of scales, each
choice leading to a different phenomenology \cite{Randall:1996zi}.

%%%%%%%%%%%%%%%%%%%%%%%%%%%%%%%%%%%%%%%%%%%%%%%%%%%%%%%%%%%%%%%%%%
%%%%%%%%%%%%%%%%%%%%%%%%%%%%%%%%%%%%%%%%%%%%%%%%%%%%%%%%%%%%%%%%%%
\section{Natural semi-direct mediation: examples}

Within the spirit outlined above, there are many ways in which one can
build concrete semi-direct gauge mediation models.

In the most conservative case, one might like to keep untouched the simple
dynamical mechanism presented in \S\,\ref{section:toy}. In this case, the
middle node has to be described by an effective SQCD with $N_f=N_c$ at scales above
$\Lambda_m$.

If we wanted just to retrofit the model of \cite{Seiberg:2008qj}, where the hidden sector is
the well known 3-2 model \cite{Affleck:1984xz}, we could
then take the messenger node $SU(N_h)$ to be the $SU(2)$ gauge group of the 3-2
model (to which we attach further to the left an $SU(3)$ node, along with
additional matter), and the visible node to be $SU(5)$
(possibly broken to the MSSM gauge group). Then, the middle node has to be taken to be $SU(7)$. Below its dynamical scale, when
$SU(7)$ confines, the model is exactly the one of
\cite{Seiberg:2008qj}, albeit with dynamically generated masses for
the messengers (and an additional adjoint in the $SU(2)$ which plays
no r\^ole at all, since it is very massive). Obviously, above the messenger
mass scale, when $SU(7)$ deconfines, we
acquire a lot more matter fields in $SU(5)$. However, if this happens
close enough to the GUT scale,
there is no Landau pole problem with having such a large number of
messengers.

Similarly, one can retrofit the semi-direct mediation model discussed in \cite{Elvang:2009gk} (the hidden sector is now
the 4-1 model \cite{Dine:1995ag,Poppitz:1995fh}), where the
messengers are charged under the abelian gauge group of the hidden sector. In this case, keeping
the visible sector to be again an $SU(5)$ GUT, we should take the middle node to be
$SU(6)$ and the rest is the same as before.

Another possibility would be not to change at all the quiver in figure \ref{quivertoy}, take $N_h=N_v=5$,
and dress the left node with a complete (hidden) $SU(5)$ family and the right node with three
($SU(5)$ GUT) families and a Higgs sector. The mediator node would then be $SU(10)$.
The hidden sector is a well known example of (incalculable) DSB model \cite{Affleck:1983vc}, with the scale
of supersymmetry breaking being of order $\Lambda_h$.
If the effective messenger mass is sufficiently above the DSB scale
$\Lambda_h$, this should be considered as a bona-fide semi-direct mediation
model, since the messengers in this case should not affect the DSB mechanism,
as in the similar cases discussed in \cite{Murayama:1995ng,Poppitz:1995fh}.
In our set up, all scales are related to dynamically
generated scales. Another very similar model would be based on taking
as the hidden DSB sector the calculable model of $SU(5)$ with two
chiral families \cite{Affleck:1984uz}.\footnote{Of course, calculability is not a value in
itself. Rather, incalculable models might even predict more generic MSSM
soft terms. In this respect, it is also worth mentioning that for the
uncalculable model of \cite{Affleck:1983vc} an exact D-brane construction (and hence, in principle,
its gravity dual description) is known in string theory \cite{Franco:2007ii}.}

Similarly, one can consider hidden sectors where supersymmetry is broken in a
metastable vacuum, as massive SQCD in the magnetic-free phase
\cite{Intriligator:2006dd}. This type of models have been considered as hidden
sectors for direct mediation of supersymmetry breaking, where the MSSM gauge
group is embedded in the flavor group of SQCD
\cite{Kitano:2006xg,Csaki:2006wi}. Here we propose to have instead
only the messengers charged under the (weakly gauged) flavor group.
The presence of an unbroken R-symmetry in the ISS supersymmetry breaking vacua
is a problem for these direct mediation models since it suppresses gaugino masses.
In our semi-direct setup the R-symmetry would result
in suppressing the off-diagonal terms in the messenger mass matrix, which in
turn also leads to a suppression of the MSSM gaugino masses. For this reason we prefer
to consider models such as the one presented in \cite{Kitano:2006xg} where, at the price of a
more complicated vacuum structure, the R-symmetry problem is overcome. In the next section,
when considering string embeddings, we review a specific semi-direct mediation model along
these lines.

We can of course continue along these lines and build many more models. In what follows, we want instead
to show how to recover models of natural semi-direct mediation within known quiver gauge theories
arising from D-branes at Calabi-Yau singularities.

%%%%%%%%%%%%%%%%%%%%%%%%%%%%%%%%%%%%%%%%%%%%%%%%%%%%%%%%%%%%%%%
%%%%%%%%%%%%%%%%%%%%%%%%%%%%%%%%%%%%%%%%%%%%%%%%%%%%%%%%%%%%%%
\section{String embeddings of retrofitted messenger sectors}
\label{string}

Though it is an interesting and perhaps amusing exercise to try and
find a model with hidden, messenger and MSSM sectors in a setup 
which could lead to a holographic gravity dual, it is not clear at
all why, for instance, we would like to embed the MSSM, which is not
strongly coupled (at least at the scales relevant to the soft terms),
in such a strict set up. Thus, we could very well be satisfied with
engineering through branes at CY singularities just the sectors which
are truly strongly coupled, like the composite messenger sector that
 we described in this note, and possibly the DSB sector. The latter could
even be taken to be an incalculable model defined by its holographic
string dual, as in \cite{Benini:2009ff}. In the same spirit, and
in the context of a full-fledged CY compactification,
one can then add visible matter by means of D7-branes \cite{Benini:2009ff}, which  live far
away from the warped throat where the hidden sector strong coupling dynamics
takes place, and are hence weakly coupled.

We present below three examples of how to embed the composite messenger
models in known quivers dual to D-branes at singularities.\footnote{Another
  approach for embedding gauge mediation in the context of D-branes at
  singularities can be found in \cite{GarciaEtxebarria:2006rw}.}
For the reasons above, we will not be interested in the details of the
visible sector.

%%%%%%%%%%%%%%%%%%%%%%%%%%%%%%%%%%%%%%%%%%%%%%%%%%%%%%%%%%%%%%%%%%%
\subsection{The orbifold of the conifold as a messenger sector}
\label{subsection:orbcon}

\begin{figure}[ht]
\centering
\vspace{-0.3cm}
\includegraphics[width=.42\textwidth]{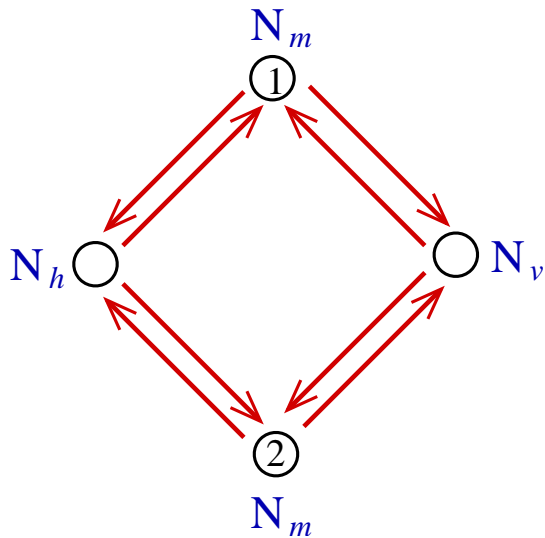}
\vspace{-0.3cm}
\caption{\small The quiver theory for a stack of (fractional) D-branes at a
$\mbox{Conifold}/\,\bZ_2$ singularity. In our realization we choose $N_m=N_h+N_v$.
\label{quiverorb}}
\end{figure}

One of the simplest string setups where to implement the mechanism discussed in \S
\ref{section:toy} is the $\bZ_2$ orbifold of the conifold, extensively used in
\cite{ABFK}. The model is just a simple variant
of our toy model: a four nodes non-chiral quiver gauge theory with
bi-fundamental matter, as depicted in figure \ref{quiverorb}.

Here we take
the gauge group to be $SU(N_h)\times SU(N_m)_1\times SU(N_v)\times SU(N_m)_2$, the
superpotential given by string theory being
\be W_\mathrm{tree}=h\left(
X_{h1}X_{1v}X_{v1}X_{1h}-X_{1v}X_{v2}X_{2v}X_{v1}
+X_{v2}X_{2h}X_{h2}X_{2v}-X_{2h}X_{h1}X_{1h}X_{h2}\right)~.
\ee
We see that instead of one mediator node, we have two. We will take
$N_m=N_h+N_v$, so that both mediator nodes are effectively like
$N_f=N_c$ SQCD. These two nodes have two separate scales
$\Lambda_1$ and $\Lambda_2$. Assuming that they both confine, we will
have an effective description in terms of two sets of messenger
fields, two sets of adjoint fields and two sets of baryons,
essentially doubling the effective fields with respect to the toy
model discussed at the beginning of this note. By solving for the
F-terms, and taking into account the constrained moduli spaces, one
can show that there are several branches. The only one where the
visible or hidden gauge groups are not higgsed is when we are on the
baryonic branch for both mediator nodes. Then, the effective superpotential
reads
\be
W_\mathrm{eff}=h\left(
M_{hv}^{(1)}M_{vh}^{(1)}-M_{vv}^{(1)}M_{vv}^{(2)}+M_{vh}^{(2)}M_{hv}^{(2)}
-M_{hh}^{(1)}M_{hh}^{(2)}\right)~.
\ee
Rescaling each meson superfield by the scale of its confined gauge
group $M_{ab}^{(i)}=\Lambda_i \Phi_{ab}^{(i)}$, we get
\be
W_\mathrm{eff}=h\Lambda_1^2 \Phi_{hv}^{(1)}\Phi_{vh}^{(1)}
-h\Lambda_1\Lambda_2
\left(\Phi_{vv}^{(1)}\Phi_{vv}^{(2)}+\Phi_{hh}^{(1)}\Phi_{hh}^{(2)}\right)
+h\Lambda_2^2 \Phi_{vh}^{(2)}\Phi_{hv}^{(2)}~.
\ee
So, if there is
some hierarchy $\Lambda_1 \ll \Lambda_2$, naturally generated by a mild
hierarchy of the UV couplings, we see that we have one pair
of messengers which are the lightest effective fields. Then come the
adjoints, and eventually we have another pair of massive
messengers which should be completely irrelevant to the physics below
the GUT scale.

This simple model can then be made phenomenologically viable by
decorating the hidden and visible nodes with the relevant matter
fields or additional nodes. The virtue of the present model is to
naturally separate the scales of the messengers from the ones of the
adjoints, making the former always the lightest. It may seem more
contrived than our simple toy model. However, from the string theory
perspective it is actually simpler, since it does not require
multitrace-like operators in order to give mass to the
adjoints.\footnote{Such operators could arise from D-branes at
orientifolded singularities.}

%%%%%%%%%%%%%%%%%%%%%%%%%%%%%%%%%%%%%%%%%%%%%%%%%%%%%%%%%%%%%%%%
\subsection{A chiral messenger sector}

\begin{figure}[ht]
\centering
\vspace{-0.1cm}
\includegraphics[width=.42\textwidth]{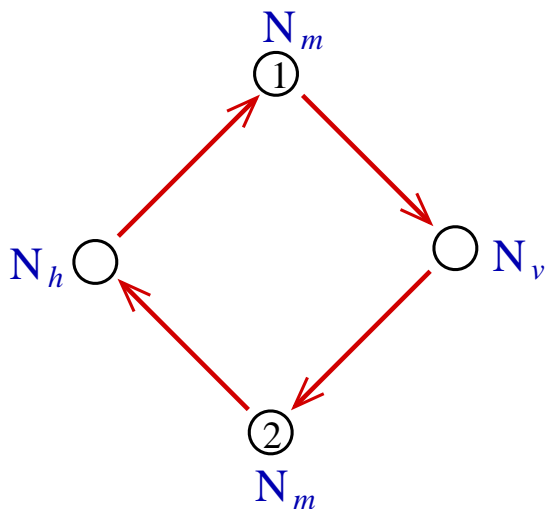}
\vspace{-0.3cm}
\caption{\small The quiver theory for a stack of (fractional) D-branes at a
$dP_3$ singularity. To avoid gauge anomalies we have $N_h = N_v$. We further require
$N_m=N_h=N_v$ for our purposes.
\label{chir}}
\end{figure}

It could be interesting to provide also an example of composite messengers
which derive from a chiral model. The added value of such a model is to
explain naturally why we only have
quartic superpotential terms. Take for instance the following model, which can
be recovered considering fractional branes at a $dP_3$ singularity (see e.g.
\cite{Argurio:2008jm}). The quiver structure is the same of the previous model, but with
half the bi-fundamental matter, see figure \ref{chir}.

The model is hence chiral
and the only superpotential one can write is
\be
W_\mathrm{tree}=h X_{h1}X_{1v}X_{v2}X_{2h}~.
\ee
Another constraint from chirality is that $N_h=N_v$, in order to prevent
gauge anomalies in the mediator nodes (let us stress again that all these conditions come out automatically
from the string construction). We further take $N_m=N_h=N_v$,
so that exactly as before when the two middle nodes confine the moduli
space splits into two, the interesting branch being the baryonic one
where the mesons are massive. The effective superpotential is then
\be
W_\mathrm{eff}=h M_{hv}M_{vh}  = h \Lambda_1 \Lambda_2 \Phi_{hv}\Phi_{vh}~.
\ee
A similar model is obtained considering D-branes at the $F_0$ singularity
(which is a different - and chiral - $\bZ_2$ orbifold of the conifold with respect to the
one considered previously), whose quiver is obtained by doubling all the bi-fundamental
fields of the $dP_3$ quiver. The only drawback in this case
is that one is left, after the middle nodes confine, with four times as many
composite messengers with the same mass.

In these models there is no need to impose symmetries that forbid the presence
of potentially dangerous lower order terms in the superpotential. Also,
there are no additional adjoints in the visible and hidden sector.
On the other hand, chirality imposes the messenger gauge
group $SU(N_h)$ to be essentially $SU(5)$, so that one might have to worry about Landau
poles, as in some direct mediation models.

%%%%%%%%%%%%%%%%%%%%%%%%%%%%%%%%%%%%%%%%%%%%%%%%%%%%%%%%%%%%%%%%
\subsection{A metastable DSB example}

In a similar fashion, one can consider string embeddings of natural semi-direct gauge
mediation where supersymmetry is broken \`a la ISS \cite{Intriligator:2006dd}. Here we aim at embedding
both the messenger and the full hidden sector in a setup derived from D-branes at singularities.

As an explicit example, we consider the model presented in \cite{Kitano:2006xg} (we refer to it as KOO). This model
breaks the R-symmetry, and therefore it does allow for gaugino mass terms when used as a model of direct gauge mediation.
As already stressed, this property is crucial also in semi-direct setups: here the gauginos are those of the messenger 
gauge group, and such a mass term is needed to provide, ultimately, masses to the MSSM gauginos.

\begin{figure}[ht]
\begin{center}
\includegraphics[width=.85\textwidth]{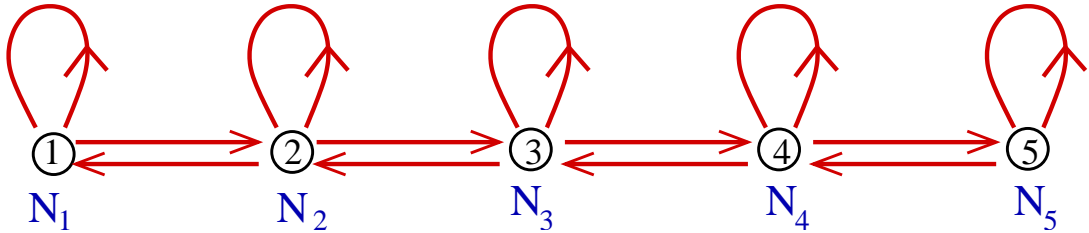}
\caption{\small The $A_5$ quiver gauge theory. The adjoint fields at each node
  are labeled $X_{ii}$ and the bi-fundamentals $q_{ij}$.
\label{a5fig}}
\label{A5quiver}
\end{center}
\end{figure}

In the KOO model the supersymmetry breaking sector and the retrofitted messenger sector arise from D-branes at a
($\mathcal{N}=2$ preserving) $A_5$ singularity, suitably deformed to a  $\mathcal{N}=1$ CY singularity
by means of a non-trivial geometric fibration (for a description of deformed $A_n$ singularities and the corresponding
field theories, see \cite{Vafa:2000wi}). The quiver gauge theory is depicted in figure \ref{a5fig}.

\noindent
The superpotential  reads
\bea
W &=& q_{12}q_{21} X_{11}-q_{21}q_{12}X_{22}+q_{23} q_{32}  X_{22}-q_{32} q_{23} X_{33} + \nonumber \\
&+&q_{34}q_{43}X_{33}-q_{43}q_{34}X_{44}+q_{45}q_{54}X_{44}-q_{54}q_{45}X_{55} + \nonumber \\
&+& \frac{M}{2} X_{11}^2+ m_1^2 X_{11} -\frac{M}{2}X_{22}^2 +
m_3^2 X_{33} + \frac{M}{2}X_{44}^2 + \frac{M}{2} X_{55}^2~,
\label{supdef}
\eea
where the last line is the $\mathcal{N}=2 \rightarrow \mathcal{N}=1$ breaking part. We have set the dimensionless
cubic coupling $h=1$ for simplicity while, as in previous examples, the mass scales $M, m_1, m_3$ can be generically
taken of the order of the Planck scale $M_p$.

Having in mind the embedding into a concrete semi-direct gauge mediation model, the rightmost node, node 5,
corresponds to the visible sector. Nodes 1 to 3 are the ISS-like sector, with node 3 becoming the
messenger gauge group, eventually. Node 4 plays then the r\^ole
of the mediator node. Let us see how things work.

It is convenient to work in the following regime for the dynamical scales of the different gauge factors
\be
\label{lambdah}
\Lambda_4>\Lambda_2>\Lambda_1\,,\,\Lambda_3\,,\,\Lambda_5~.
\ee
This hierarchy defines the steps one has to perform
in order to obtain the low energy theory.\footnote{This is essentially for
  convenience. We could choose not to enforce a hierarchy between 
$\Lambda_2$ and $\Lambda_4$ (while keeping them larger than the other scales),
the analysis would be slightly more involved but the results unchanged.}

Given that the mass parameters entering the superpotential (\ref{supdef}) are of the order of the highest scale in the problem,
$M_p$, we first have to integrate out the massive fields $X_{11},X_{22},X_{44},X_{55}$, obtaining
\bea
W_{\mathrm{eff}}&=&-\frac{1}{M}q_{12}q_{23}q_{32}q_{21}+\frac{1}{2 M} (q_{23} q_{32})^2
-\frac{1}{2M} (q_{34}q_{43})^2+\frac{1}{M}q_{34}q_{45}q_{54}q_{43}
-\frac{1}{M}(q_{54}q_{45})^2\non \\
&&-q_{32}q_{23}X_{33}+X_{33}q_{34}q_{43}
-\frac{m_1^2}{M}q_{12}q_{21}+ m_3^2 X_{33}~.
\eea
Then, flowing to the IR, node $4$ develops strong dynamics. To reproduce the simple dynamics of our toy model,
we choose the ranks such that $N_3+N_5=N_4$. This way node $4$ has $N_f=N_c$ and undergoes confinement
with a modified moduli space. Selecting the baryonic branch, as before, and integrating out the very massive fields
one obtains
\be
\label{super3}
W_{\mathrm{eff}}=
-\frac{1}{M}q_{12}q_{23}q_{32}q_{21}
-\frac{m_1^2}{M}q_{12}q_{21}+
\frac{m_3^2}{M}q_{32}q_{23}
+\frac{1}{M} N_{35}N_{53}~,
\ee
where $N_{35} = q_{34} q_{45}$ and $N_{53} =  q_{54} q_{43}$ are the mesons associated to node 4.
The first three terms in eq.~(\ref{super3}) correspond to the electric phase of the
KOO model, which breaks supersymmetry into a metastable vacuum. The messenger fields
$N_{35}$ and $N_{53}$ couple to it only through the gauge group at node $3$.

We now proceed to obtain the low energy description. According to the hierarchy
among the scales we chose, eq.~(\ref{lambdah}), node $2$ becomes then strongly coupled, and we have to dualize it.
To recover the KOO model, we choose the ranks $N_1=N_f-N_c$, $N_2=N_3=N_c$.
In the dual description, node $2$ is a $SU(N_f-N_c)$ gauge group, and in terms of magnetic
dual variables the superpotential reads
\bea
W_{\mathrm{eff}}&=&-\frac{m_1^2}{M}M_{11}+\frac{m_3^2}{M}M_{33}+ \frac{1}{\Lambda_2} \left(M_{11} t_{12} t_{21}+
M_{33}t_{32}t_{23}+M_{13}t_{32}t_{21}+M_{31}t_{12}t_{23} \right) \non\\
\phantom{W_{\mathrm{eff}}}& &-\frac{1}{M} M_{13}M_{31}+\frac{1}{M} N_{35}N_{53}~,
\eea
where $t_{ij}$ are the magnetic quarks. In order to estimate the scales of the theory we give the mesons
canonical dimensions, inserting the proper dynamical scales, finally obtaining
\be
W_{\mathrm{eff}}= - \bar m^2 M_{11}+\bar\mu^2 M_{33}+M_{11} t_{12} t_{21}+
M_{33}t_{32}t_{23}
+M_{13}t_{32}t_{21}+M_{31}t_{12}t_{23}
- m_z M_{13}M_{31} + m N_{35}N_{53}~,
\ee
where
\be
\label{parameters}
\bar m^2 =\frac{m_1^2 \Lambda_2}{M} \quad
\bar \mu^2 =\frac{m_3^2 \Lambda_2}{M}
\quad m_z= \frac{\Lambda_2^2}{M}\quad m=\frac{\Lambda_4^2}{M}~,
\ee
and recall that $M, m_1, m_3 = {\cal O}(M_p)$.
As discussed in \cite{Kitano:2006xg}, in order for the metastable ISS-like vacua to have a
sufficiently large lifetime we have to require $\bar\mu < \bar m$ and $m_z < \bar m $. The first inequality
is easily accomplished by a mild tuning of the masses $m_1,m_3$ which, generically, are of the
same order of magnitude. The second is easily seen to hold (strongly) since
we assume $\Lambda_2 \ll M_p$. Similarly, we also have $m\ll \bar m$.

The relative hierarchy between $m$ and $m_z$ depends on whether we implement
the hierarchy (\ref{lambdah}) or not.  Roughly, what we need is 
the messenger supersymmetric mass $m$ to lie
 within the mass of the gauginos associated to
 node 3 (which depends on $m_z$ as we review below) and $\bar m$, the highest mass 
 scale in the hidden sector, as discussed in \cite{Randall:1996zi}.

As anticipated we have realized a model of natural (metastable) semi-direct mediation. In our setup the MSSM is
at node $5$. It is not directly coupled to the supersymmetry breaking sector,
but it is coupled to the massive fields $N_{35}$ and $N_{53}$, which are
the composite messengers. The mediation is semi-direct, since the messengers are coupled
to the susy breaking sector only through gauge interactions (node $3$, the messenger gauge group, which in this
case enters just as a flavor group within the DSB mechanism).

%%%%%%%%%%%%%%%%%%%%%%%%%%%%%%%%%%%%%%%%%%%%%%%%%%%%%%%%%%%%%
\subsubsection*{\it Computing the messenger susy-breaking spectrum}

Since this model has an explicit (and calculable) hidden sector, we can go a little
further.

We would like to compute the mass splitting for the messengers $N_{35}$ and $N_{53}$.
In a semi-direct scenario as the one discussed here, the messengers feel the breaking of supersymmetry
exactly as the matter chiral fields of the MSSM in a model of gauge mediation. The only difference here is
that they already have a mass at tree level $m$. This implies that, besides
ordinary diagonal mass terms (which contribute to the supertrace), off-diagonal
masses are also generated at two loops. The messenger masses are
easily computed using standard techniques.

The diagonal supersymmetry breaking
mass for the messengers are as the soft scalar masses computed
in \cite{Kitano:2006xg}
\be
m_{d}^2%=m_{sc}^2
\sim
\alpha_3^2 \bar N \left(\frac{\bar \mu^2}{\bar m}\right)^2~,
\label{mdmess}
\ee
where $\alpha_3=g_3^2/4\pi$ is the loop factor of the gauge group at node 3
and $\bar N=N_1+N_3 -N_2$.

The off-diagonal masses for the messengers can also be
computed. The computation involves the
supersymmetric mass of the messengers $m$
and the non-supersymmetric mass of the gauginos of node $3$,
which is \cite{Kitano:2006xg}
\be
\label{mgaugino}
m_{\lambda} \sim \alpha_3 \bar N  \frac{\bar \mu^2 m_z}{\bar m^2}~.
\ee
Henceforth, we will take $\bar N={\cal O}(1)$ and drop it from the expressions.
To perform the computation we parameterize the gaugino mass as in the
general gauge mediation (GGM) formalism \cite{Meade:2008wd}
\be
\label{mgaugino2}
m_\lambda \sim \alpha_3 M B_{1/2}(0)~,
\ee
where
$B_{1/2}(p^2/M^2)$
is a function characterizing a
current-current correlator of the
hidden sector.
Since we are only interested in order of magnitude estimates, we approximate
it as a step 
function, i.e. $B(x)=1$ if $x\in (0,1)$, $B(x)=0$ otherwise. $M$ is the scale
emerging from the KOO sector.
It can be obtained comparing eqs.~(\ref{mgaugino}) and (\ref{mgaugino2}), and reads
$M=\bar \mu^2 m_z/\bar m^2$.

The off-diagonal mass for the messengers is then given by a
diagram involving the massive messenger fermion and the
hidden gauginos of node $3$
\be
m_{\mathit{off}}^2 \sim \alpha_3^2\int d^4 p \, \frac{1}{p^2} \frac{m}{p^2+m^2} \, M B_{1/2}(\frac{p^2}{M^2})~.
\ee
It can be evaluated using the approximation we explained before for the function $B_{1/2}(x)$ giving
\be
m_{\mathit{off}}^2\sim 
 \alpha_3^2 m \frac{\bar \mu^2 m_z}{\bar m^2} \log(1+\frac{\bar \mu^4 m_z^2}{\bar m^4 m^2})~.
\label{moff}
 \ee
We see that the off-diagonal mass, which is going to be the leading
contribution to the visible gaugino mass, peaks when $m$ is of the same
order as the mass of the hidden gaugino divided by the coupling constant,
$m_{\lambda}/\alpha_3$, see eq.~\eqref{mgaugino}.

If we work in the regime (\ref{lambdah}), the ratio $m_\lambda/(\alpha_3 m)$ is
easily suppressed by a few orders of magnitude and $m_{\mathit{off}}^2$
can be approximated as
\be
m_{\mathit{off}}^2 \sim  \frac{ \alpha_3^2}{m}
\left(\frac{\bar \mu^2 m_{z}}{\bar m^2} \right)^3~.
\ee
Hence it results suppressed in the supersymmetry breaking
scale set by the gaugino mass (\ref{mgaugino}).

We could think however of a different scenario.
In order to make  $m_\lambda/( \alpha_3 m) = {\cal O}(1)$
we take a different hierarchy with respect to the one in eq.~(\ref{lambdah}),
namely  $\Lambda_2 > \Lambda_4$. We have
\be
\frac{m_\lambda}{\alpha_3 m} \sim  \frac{\bar \mu^2}{\bar m^2} \frac{m_z}{m}
\sim  \frac{m_3^2}{m_1^2} \frac{\Lambda_2^2}{\Lambda_4^2}~.
\ee
Recalling what we said after eq.~\eqref{parameters}, the ratio $m_3/m_1$ in the expression above is necessarily smaller than
one; but we see that even with a small hierarchy like $\Lambda_2\sim 10
\Lambda_4$ we can achieve a not too small ratio $m_\lambda/(\alpha_3 m)$ so that
the $\log$ in eq.~(\ref{moff}) is roughly of order one.

Thus, in this regime,
we have potentially unsuppressed visible gaugino masses, proportional to
\be
m_{\lambda,\mathrm{vis}} \sim  \alpha_\mathrm{vis} \frac{m_{\mathit{off}}^2}{m}
\sim  \alpha_\mathrm{vis} \alpha_3 m_\lambda
\sim \alpha_\mathrm{vis} \alpha_3^2 m ~. \label{vis3loop}
\ee
As in \cite{Randall:1996zi}, the visible gaugino mass turns out to be a
three-loop effect (one MSSM loop times two hidden loops). Assuming the
messenger gauge group at node 3 to be rather weak, this leads to a messenger
mass which is $10^9-10^{11}$ GeV, which seems phenomenologically acceptable.
 
The question is of course what is the ratio of this gaugino mass to the MSSM
sfermion masses. If we were in a set up similar to minimal gauge mediation, 
we could immediately see that this ratio would be of order one. However, we are
not in such a set up, since we have a sizable contribution from the diagonal
messenger masses (\ref{mdmess}) to the supertrace over the messenger sector.
As discussed in \cite{Poppitz:1996xw}, in such a situation the sfermion masses
become model dependent. This is expected, since by turning on
the supertrace we cover more parameter space. Hence, in order to discuss in
more detail the MSSM soft spectrum one should set up the computation along
the lines of \cite{Poppitz:1996xw}.\footnote{
Note that, as
  discussed in  \cite{Randall:1996zi,Poppitz:1996xw}, a positive supertrace
  over the messenger sector would result in
  tachyonic masses for the MSSM sfermions, which is of course
  phenomenologically ruled out.} 
This is however beyond the scope of the
present paper. What we have shown is that in this model 
the MSSM gaugino masses are not a priori suppressed
by higher orders in the off-diagonal
messenger masses, at least in some range of parameters.
This suggests that F-term
suppression of gaugino masses is not generic in semi-direct gauge
mediation.

%%%%%%%%%%%%%%%%%%%%%%%%%%%%%%%%%%%%%%%%%%%%%%%%%%%%%%%%
%%%%%%%%%%%%%%%%%%%%%%%%%%%%%%%%%%%%%%%%%%%%%%%%%%%%%%%%

\section{Conclusions and outlook}

We have shown that semi-direct gauge mediation can be easily made natural, by dynamically generating supersymmetric masses
for the messenger fields. The simple mechanism we have proposed
turned out to be quite generic in string theory when realized by D-brane embeddings.

We have discussed a few explicit examples by considering D-branes at chiral
and non-chiral CY singularities. The ease in which models of semi-direct mediation arise suggests that a
large portion of their parameter space can be covered by such
constructions. As an example of such possibility, we have shown that in a calculable model where supersymmetry
is broken in a metastable vacuum and the messenger gauge group is a (weakly gauged) flavor group of the hidden sector,  one
may choose a regime where the visible gaugino masses are not necessarily suppressed. This regime is not easily
achieved in models of semi-direct gauge mediation where the messenger gauge group is a
genuine gauge group of the hidden sector, and is higgsed \cite{Seiberg:2008qj,Elvang:2009gk}.
Conversely, the generation of D-term-like contribution to the diagonal scalar masses for the messengers, induced
by the supersymmetry breaking F-terms, seems specific to the latter class of higgsed models.

These facts suggest that, from a phenomenological point of view, besides generic predictions encompassed
by any model of semi-direct gauge mediation, there are others which seem to hold only for some specific subclass
of models. It would  be desirable to analyze further
the phenomenology of semi-direct gauge mediation. In this respect, recasting semi-direct gauge mediation
in the formalism of general gauge mediation \cite{Meade:2008wd} might be helpful. Efforts in this direction are under
way.

\vskip20pt
\centerline{\bf \large Note Added}

Continuing our investigation of models of semidirect gauge mediation, we found that also our model does not escape the mechanism of suppression of visible gaugino masses presented in~\cite{AGLR}. We showed this explicitly at the diagrammatic level in~\cite{generalfinal}. Formula (\ref{vis3loop}) above is thus unfortunately incorrect as it stands and the first non zero contribution to the visible gaugino mass is expected at higher loop order.

\vskip20pt
\centerline{\bf \large Acknowledgments}

We thank Ken Intriligator for useful discussions, and Shamit Kachru and Brian Wecht for thoughtful
comments on a preliminary version of this paper. R.A. and M.B. would like to thank the organizers of the workshop
``SUSY Breaking 09'' held at Durham for arranging a very stimulating meeting, and its participants for useful discussions
and exchanges of ideas. 
The research of R.A. is supported in part by IISN-Belgium (conventions
4.4511.06, 4.4505.86 and 4.4514.08). R.A. is a 
Research Associate of the Fonds
de la Recherche Scientifique--F.N.R.S. (Belgium).
The research of G.F. is supported in part by the Swedish Research Council (Vetenskapsr{\aa}det)
contract 621-2006-3337. Contribution from the L\"angmanska Kulturfonden and the Wilhelm och Martina Lundgrens
Veteskapsfond are also gratefully acknowledged.
A.M. is
supported in part by 
FWO-Vlaanderen through project G.0428.06. R.A. and A.M. are also supported in part
by the Belgian Federal Science Policy Office 
through the Interuniversity 
Attraction Pole IAP VI/11.

\end{document}